\documentclass[aps,pra,twocolumn,superscriptaddress,longbibliography]{revtex4-2}

\pdfoutput=1
\usepackage[english]{babel}
\usepackage{xcolor}
\newcommand{\comment}[1]{}
\usepackage{amsmath}
\usepackage{amsfonts}
\usepackage{tensor}
\usepackage{dsfont}
\usepackage{physics}
\usepackage{graphicx}
\usepackage[colorlinks=true, allcolors=blue]{hyperref}
\usepackage{siunitx}

\begin{document}

\title{Pulse engineering via projection of response functions at infinite nonlinear order}
\author{Lia Kley}
\email{lia.kley@uni-hamburg.de}
\affiliation{Zentrum f\"ur Optische Quantentechnologien, Universit\"at Hamburg, 22761 Hamburg, Germany}
\affiliation{Institut für Quantenphysik, Universit\"at Hamburg, 22761 Hamburg, Germany}
\affiliation{The Hamburg Centre for Ultrafast Imaging, 22761 Hamburg, Germany}
\author{Ludwig Mathey}
\affiliation{Zentrum f\"ur Optische Quantentechnologien, Universit\"at Hamburg, 22761 Hamburg, Germany}
\affiliation{Institut für Quantenphysik, Universit\"at Hamburg, 22761 Hamburg, Germany}
\affiliation{The Hamburg Centre for Ultrafast Imaging, 22761 Hamburg, Germany}

\begin{abstract}
Optimal control problems arise in a wide range of scientific disciplines, but the corresponding optimization algorithms often display a strong dependence on hyperparameters that significantly influence performance and convergence. For the optimal implementation of quantum algorithms, these challenges are further amplified by high-dimensional control landscapes and the need for high-fidelity operations. Here, we propose an algorithm for optimal control problems in quantum computing to efficiently generate high-fidelity control protocols for multi-qubit systems in a hyperparameter and gradient free manner. The method, referred to as Pulse Engineering via Projection of response functions at infinite nonlinear order (PEPRino), leverages the framework of response theory to navigate the control landscape to find high-fidelity implementations. This is achieved by determining the control landscape via response functions to infinite order, efficiently evaluated by resummation in terms of the first and second order response function. To demonstrate the approach, we apply it to quantum systems consisting of two and three qubits for the optimal implementation of the Quantum Fourier Transform (QFT). We benchmark the proposed algorithm against the Chopped Random Basis (CRAB) algorithm utilizing the Nelder-Mead method, focusing on the 2-qubit scenario. The results indicate faster convergence regarding iteration steps and computational time, highlighting the advantages of our approach.

\end{abstract}
\maketitle

\section{Introduction}
\label{sec:introduction}
\begin{figure}
    \centering
    \includegraphics{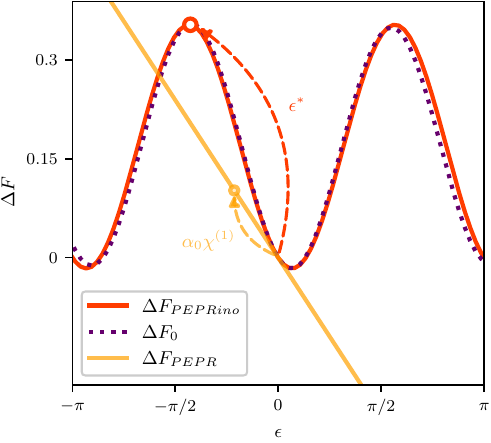}
    \caption{\textbf{Fidelity landscape and update strategy.} We show the fidelity landscape $\Delta F_0$ as a function of the parameter $\epsilon$, corresponding to a shift of a trainable parameter $\theta_{j,k}$ of a control operator $B_j$, for a two-qubit system discussed in the text. Specifically, the shift is given by ${\theta_{j,k} \rightarrow \theta_{j,k} - \frac{2\epsilon}{t_f}\sin\left(\frac{\pi k t_r}{t_f}\right)}$ for a time $t_r$ in $[t_0, t_f]$. The resulting fidelity changes are represented by the dotted purple line. Additionally, we show ${\Delta  F_{PEPRino}  =\frac{1}{2} \chi^{(1)} \sin(2\epsilon) + \frac{1}{4} \chi^{(2)} \left(1 - \cos(2\epsilon) \right)}$, which is the approximated fidelity landscape computed from the first and second order susceptibility of the system (red line). This approximation is used in the PEPRino method. The optimal value of $\epsilon$ is $\epsilon^*$, which is the parameter maximizing $\Delta F_{PEPRino}$ with the smallest absolute value, and used in the PEPRino parameter update ${\theta_{j,k} \rightarrow \theta_{j,k} - \frac{2\epsilon^*}{t_f}\sin\left(\frac{\pi k t_r}{t_f}\right)}$. Lastly, we show the fidelity approximation ${ \Delta  F_{PEPR}}=\epsilon \chi^{(1)}$ used in PEPR (orange line), which is a linear approximation of the landscape, where the parameter update ${\theta_{j,k} \rightarrow \theta_{j,k} - \frac{2\alpha_0}{t_f}\sin\left(\frac{\pi k t_r}{t_f}\right)\chi^{(1)}}$ depends on a learning rate $\alpha_0$ and the linear response, but not on higher orders.}
    \label{fig:Fig1}
\end{figure}

Optimal control plays a central role in a wide range of applications, from classical engineering to quantum technologies~\cite{doria2011optimal,li2017hybrid,koch2022quantum,chen2025robust}. In the context of quantum information processing, it provides a powerful framework for designing high-fidelity operations in the presence of physical constraints and noise~\cite{PhysRevA.111.052625,10828532,czischek24}. In particular, the implementation of quantum gates with high precision is essential for scalable quantum computation on various platforms~\cite{evered2023high,PhysRevX.15.011009,PhysRevLett.133.013401,zhou2025high,bartling2025universal,loschnauer2025scalable}. However, the control landscape typically becomes increasingly complex with system size. This makes efficient optimization a challenging task~\cite{mcclean2018barren,ge2022optimization}. Various approaches have been developed to address this problem. Variational Quantum Algorithms (VQAs) have emerged as a hybrid quantum-classical approach to quantum control~\cite{cerezo2021variational,PhysRevA.92.042303}. In a VQA, a parameterized quantum circuit is iteratively optimized using classical algorithms to minimize a cost function, such as gate infidelity. These algorithms are particularly suited for noisy intermediate-scale quantum (NISQ) devices~\cite{RevModPhys.94.015004,preskill2018quantum}. Traditional approaches include gradient-based methods~\cite{khaneja2005optimal,PhysRevResearch.7.013101} as well as gradient-free techniques such as CRAB and the downhill simplex Nelder-Mead method~\cite{PhysRevA.84.022326,10.1093/comjnl/7.4.308}. 

In this paper, we present a pulse engineering method which we refer to as pulse engineering via projection of response functions at infinite nonlinear order (PEPRino). Our method extends the approach of pulse engineering via projection of linear response  functions~\cite{PhysRevResearch.7.013101,kley2024optimalrecoilfreestatepreparation} by considering response functions at infinite nonlinear order. In our approach, we apply time-local perturbations and evaluate the response of the system, with the aim of generating optimal implementations of desired operations, resulting in optimal fidelity. We consider a Hamiltonian that depends on a set of trainable parameters which is the search space in which an optimal implementation is to be found. The trainable parameters are the prefactors of a set of mode functions, for which we choose sine functions, resulting in a temporally non-local parametrization~\cite{broers2024mitigated}. The linear combination of trainable parameters and mode functions are the control functions. Each of these control functions is associated with a control operator, which, taken together, form the trainable part of the Hamiltonian. In each update step, we consider a small perturbation of one of the control operators and determine the resulting response of the fidelity at infinite nonlinear order. By projection onto the control functions, the multi-parameter update is generated. We show the fidelity landscape and update strategy in Fig.~\ref{fig:Fig1}. The optimization is done in a completely hyperparameter-free manner, which is a key advantage over other optimal control techniques, as it simplifies the usability and improves the applicability.

By removing the need for hyperparameter tuning, our algorithm simplifies the control process, making high-fidelity quantum operations more accessible for experimental implementations and scalable to complex systems ~\cite{brif2010control}. It is essential for overcoming current limitations in quantum technology by advancing quantum control and paving the way for more robust, efficient, and user-friendly quantum systems~\cite{glaser2015training,preskill2018quantum}.

This paper is structured as follows. In Sect.~\ref{sec:model}, we introduce the PEPRino method and motivate the underlying model. In Sect.~\ref{sect:2}, we demonstrate our main example to present this method, which is the optimization of the Quantum Fourier transform (QFT)~\cite{RevModPhys.68.733} on a $N$-qubit system driven by Rabi pulses and coupled via Ising interactions. We focus on the case of two qubits, with the results being discussed in Sect.~\ref{sec:high-fidelity-operations}, where we compare the optimization using PEPRino with the use of the CRAB algorithm, specifically employing the Nelder-Mead method. While there exist various implementations of this optimization technique, in the following discussion we refer to this particular implementation simply as CRAB. Additionally, we show our results for the optimization in the case of three qubits using PEPRino. In Sect.~\ref{sec:conclusion}, we conclude our findings.

\section{Method}
\label{sec:model}
We propose a pulse-engineering method based on the response to time-local perturbations with the objective of finding optimal implementations of quantum operations. This ansatz extends the previously formulated method of PEPR~\cite{PhysRevResearch.7.013101} to infinite nonlinear order, leading to a gradient- and hyperparameter-free setup. We consider a Hamiltonian
\begin{align}
    H(t) = H_0+H_\theta(t), 
    \label{eq:fullHamiltonian}
\end{align}
where $H_\theta(t)$ is the trainable part of the Hamiltonian, and $H_0$ is the untrainable part. In the example below, we have $H_0 =0$. $H_\theta(t)$ depends on a set of trainable parameters $\theta_k$, with $k=1,...,N_\theta$. $H_\theta(t)$ is of the form
\begin{align}
    H_\theta(t) = \sum_{j=1}^{n_j} \theta_j(t)B_j
    \label{eq:Htheta}
\end{align}
with the control operators $B_j$, with $j = 1,...,n_j$.
Each control operator $B_j$ is associated with a subset of the parameters $\theta_k$ via the control function $\theta_j(t)$, specifically 
\begin{align}
    \theta_j(t)=\sum_{k=1}^{n_m}\theta_{j,k}f_{j,k}(t)
    \label{eq:thetaparam}
\end{align}
with the $n_m$ mode functions $f_{j,k}(t)$. Next, we introduce a time-dependent perturbation on the control operator $B$ of the form 
\begin{align}
    H_{\text{pr}}(t) = -F(t)B.
    \label{eq:Hpertgeneral}
\end{align}
In the following, we discuss the response of the system to this perturbation. The general expression for the expectation value of an observable $A$ at a time $t$ is
\begin{align}
    \langle A(t)\rangle &=\Tr\left(\rho(t)A\right)=\Tr\left(U(t,t_0)\rho(t_0)U^\dagger (t,t_0)A\right).
    \label{eq:expectation_A_ansatz}
\end{align}
Here, $U(t,t_0)$ denotes the time evolution operator and $\rho(t_0)$ is the initial state of the system. In the following, we work in the interaction picture. To 0-th order, this recovers the expectation value of $A$ in the unperturbed system, i.e.
\begin{align}
    \langle A(t)\rangle_0 = \Tr(\rho(t_0)A(t)) = \langle A \rangle_0.
\end{align}
The change of the expectation value of the observable due to the perturbation is
\begin{align}
    \Delta \langle A(t)\rangle &= \langle A(t)\rangle -\langle A \rangle_0.\label{eq:integralexpression_expectation_A}
\end{align}
Thus, the induced change to the expectation value of $A$ at n-th order reads
\begin{align}
    \Delta \langle A(t)\rangle_n &=\frac{1}{n!}\frac{1}{(i\hbar)^n} \int_{t_0}^t   dt'\ldots dt'^{n}\langle[\ldots\nonumber\\
    \quad\ldots[[&A(t),H_\text{pr,I}(t')],H_\text{pr,I}(t'')],\ldots,H_\text{pr,I}(t'^{n})]_n\rangle_0 .
\end{align}
Now, we consider the perturbation to be time-local at a random time $t_r$, i.e.
\begin{align}
    H_\text{pr,I}(t)=-\epsilon\delta(t-t_r)B,
    \label{eq:Hperturbation}
\end{align}
i.e. $F(t) = \epsilon \delta(t-t_r)$. Then the change of the expectation value of the observable $A$ simplifies to
\begin{align}
    \Delta \langle A(t)\rangle&=\sum_n\frac{(-\epsilon)^n}{n!}\frac{1}{(i\hbar)^n} \langle[\ldots\nonumber\\
    &\quad\ldots[[A(t),B(t_r)],B(t_r)],\ldots,B(t_r)]\rangle_0 . \label{eq:final_expression_DeltaA}
\end{align}
To determine the change of the fidelity of the implementation, we set the observable $A$ to the target state $\rho_\text{target}=V\rho(t_0)V^\dagger$, associated with the desired target transformation $V$. We find that the change in the expectation value, Eq.~(\ref{eq:final_expression_DeltaA}), corresponds to the change in fidelity $\Delta F$
\begin{align}
    \Delta \langle& A(t)\rangle=\Delta  F =\sum_n\frac{\epsilon^n}{n!}\chi^{(n)}(t_r) \label{eq:fid_as_series}
\end{align}
with the susceptibilities
 \begin{align}
\chi^{(n)}(t_{r}) &= \left(\frac{i}{\hbar}\right)^n Tr\Big(\rho_\text{target}U(t_f,t_r)[B,[\ldots\nonumber\\
    &\quad\ldots,[B,U(t_r)\rho(t_0)U^\dagger(t_r)]\ldots]_nU^\dagger(t_f,t_r)\Big).\label{eq:chi_series}
\end{align}
Next, we consider a system of qubits, and control operators consisting of Pauli matrices, i.e. $B_j\in\{\vec{\sigma}\}$, or tensor products of Pauli matrices. Then, Eq.~(\ref{eq:fid_as_series}) can be simplified further. We find that due to the $SU(2)$ algebra of the Pauli matrices, the sequence of nested commutators alternates between two linearly independent operators. Even-order terms are proportional to the second-order commutator, and odd-order terms to the first-order commutator:
\begin{align}
    i^{2n+1}\left[B_j,\cdots\left[B_j,\rho(t)\right]\cdots \right]_{2n+1}= (-4)^n\cdot i\left[B_j,\rho(t)\right] \label{eq:nestedPaulicommutatorodd}
\end{align}
with $n>0$, and
\begin{align}
i^{2n}\left[B_j,\cdots\left[B_j,\rho(t)\right]\cdots \right]_{2n} = (-4)^{n-1}\cdot i^2\left[B_j,\left[B_j,\rho(t)\right]\right]\label{eq:nestedPaulicommutatoreven}
\end{align}
with $n>1$. A detailed calculation is presented in App.~\ref{app:sec-Paulimatrices}. Consequently, out of the series of susceptibilities in Eq.~(\ref{eq:chi_series}) to infinite order, only the first and second-order susceptibilities, $\chi^{(1)}$ and $\chi^{(2)}$, need to be calculated explicitly. We can resum the series for the change in fidelity \(\Delta  F\) in Eq.~(\ref{eq:fid_as_series}) as
\begin{align}
\Delta  F_{PEPRino}  &=\frac{1}{2} \chi^{(1)} \sin(2\epsilon) + \frac{1}{4} \chi^{(2)} \left(1 - \cos(2\epsilon) \right).\label{eq:explicit_series_DeltaF}
\end{align}
A more detailed derivation is found in App.~\ref{app:sec-derivation-series}.
The values of \(\epsilon\) that maximize or minimize $\Delta F_{PEPRino}$, are:
\begin{equation}
    \epsilon_n = \frac{1}{2} \arctan\left(\frac{-2\chi^{(1)} }{\chi^{(2)}}\right) + \frac{n\pi}{2}, \quad n \in \mathbb{Z}.
\end{equation}
We evaluate the second derivative to classify the extrema, and select the \(\epsilon\) with negative second derivative, i.e., a maximum in $\Delta F_{PEPRino}$, that has the smallest absolute value, denoted as $\epsilon^*$. This approximation of the fidelity landscape and our update strategy are visualized in red in Fig.~\ref{fig:Fig1}. Following the update strategy of PEPR~\cite{PhysRevResearch.7.013101}, represented in orange in Fig.~\ref{fig:Fig1}, we parameterize the control pulses as a finite set of sine modes~\cite{broers2024mitigated}, as we described above. We update the trainable parameters according to
\begin{align}
\theta_{j,k}\rightarrow \theta_{j,k}-\frac{2\epsilon^* }{t_f}\sin\left(\frac{\pi kt_r}{t_f}\right)
\end{align} 
for the parameters $\theta_{j,k}$, associated with the operator $B_j$. The factor $2/t_f\sin\left(\pi k t_r/t_f\right)$ in the update form arises from projecting the $\delta$-function considered in the perturbation Eq.~\eqref{eq:Hperturbation} onto the finite set of mode series in the control pulse parameterization~\cite{PhysRevResearch.7.013101}. Additionally, we show the actual change in fidelity resulting from updates of this form in Fig.~\ref{fig:Fig1} as the dotted purple line $\Delta F_0$ as a function of $\epsilon$. We note that the expression given in Eq.~\eqref{eq:explicit_series_DeltaF} provides a good estimate of the fidelity landscape, including capturing the nearby maxima and minima. $\Delta F_0$ shows small deviations from $\Delta F_{PEPRino}$ for larger values of $\epsilon$, in this example for $|\epsilon| \geq \pi/2 $. However, these deviations are not critical for the performance of the method, and diminish with an increasing number of modes considered in the parameterization.

\section{Model}
\label{sect:2}
We compare our method to the established method of the chopped random basis (CRAB) algorithm~\cite{PhysRevA.84.022326}, in which the optimization is performed using the direct search method of the Nelder-Mead simplex algorithm~\cite{10.1093/comjnl/7.4.308, doria2011optimal, lagarias1998convergence}, and refer to this implementation as the CRAB algorithm, for simplicity. We investigate the performance of both methods in learning an implementation of the quantum Fourier transform (QFT). Throughout this work we use the Ising Hamiltonian~\cite{stinchcombe1973ising}, given by 
\begin{equation}
    H(t)=\sum_{j=1}^{n_q}\left(h_x^j(t)\sigma_x^j+h_y^j(t)\sigma_y^j\right) + \sum_{j=1}^{n_q-1}J^j(t)\sigma_z^j\sigma_z^{j+1},
\end{equation}
where $h_x^j(t), h_y^j(t)$ and $J^j(t)$ are time-dependent controllable parameters. We note that it might be desirable to restrict the magnitude of these control functions, by a maximal value for the Rabi frequency or the Ising coupling, for example, see~\cite{PhysRevResearch.7.013101}. We note that for the optimization examples given below, we obtain converged high-fidelity implementations with finite, non-diverging values for the control functions, while for other examples constraints might be necessary. For illustration, the control operators for $n_q=2$ atoms in our model are 
\begin{equation}
    \vec{B} = \{\sigma_x\otimes\mathds{1} , \mathds{1} \otimes\sigma_x,\sigma_y\otimes\mathds{1} , \mathds{1} \otimes\sigma_y,\sigma_z\otimes\sigma_z\}. \label{eq:2qubitoperators}
\end{equation} 
The corresponding control pulses are parameterized in a time-nonlocal manner~\cite{broers2024mitigated}, i.e.
\begin{align}
h_{x/y}^j(t)&= \sum_{k=1}^{n_m}\theta_{x/y,j,k}\sin\left(k\pi t/t_f\right) \\
    J^j(t)&= \sum_{k=1}^{n_m}\theta_{J,j,k}\sin\left(k\pi t/t_f\right) .
\end{align}
In total, this gives $N_\theta = (3n_q-1)n_m$ trainable parameters, in particular, the set of parameters composed of $\theta_{x/y,j,k}$ and $\theta_{J,j,k}$ make up the set $\theta_k$. For the optimal implementation of the QFT~\cite{weinstein2001implementation}, represented in a matrix $V$ with matrix elements~\cite{PhysRevResearch.7.013101}
\begin{equation}
    V_{s,t}=2^{-n_q/2}\exp\{i2\pi st2^{-n_q}\},
\end{equation}
we randomly sample the initial density matrix $\rho(t_0)$ as a tensor product of single-qubit pure states. For each qubit, a random three-dimensional real vector is drawn from a normal distribution and normalized, defining a pure state on the Bloch sphere. The corresponding single-qubit density matrices are then constructed and combined via tensor products to form the full $N$-qubit state $\rho(t_0)$. Subsequently, we randomly choose a control operator $B_j\in \vec{B}$ as well as a random time $t_r\in[t_0,t_f]$, where $t_f$ is the transformation time. Finally, $\chi^{(1)}$ and $\chi^{(2)}$  are extracted following Eq.~(\ref{eq:chi_series}) with the target state $\rho_\text{target}= \text{QFT} \rho(t_0)\text{QFT}^{\dagger}$. As a measure of the optimality of the current implementation, we evaluate the infidelity of an ensemble $n_E$ of initial states $\rho_i(t_0)$ compared to the corresponding target states
\begin{equation}
    \langle 1-F\rangle =1-\frac{1}{n_E}\sum_\text{i=1}^{n_E}\text{Tr}(\rho_\text{i,target}^\dagger \rho_\text{i}(t_f)).
    \label{eq:1-F}
\end{equation}
The target states are obtained after time propagation with updated trainable parameters, which are updated following
\begin{align}
\theta_{j,k}\rightarrow \theta_{j,k}-\frac{2\epsilon }{t_f}\sin\left(\frac{\pi kt_r}{t_f}\right),
\end{align} 
where $\epsilon$ is determined following the proposed strategy, see Sec.~\ref{sec:model}. In addition to the sampling of the initial state via tensor products of random single-qubit states, we also consider sampling via 2-design for the initial state, or a large random sample of a 2-design, see App.~\ref{app:sec-2design}. For PEPRino, these evaluation methods are used for visualization purposes only. In this context, they represent alternative approaches for the evaluation of the infidelity, with neither offering a significant advantage in this example. In contrast, the CRAB ansatz requires a faithful approximation of the full average over the state space for the algorithm to function properly. In practice, using only a subset of a 2-design or an ensemble of randomly initialized states introduces noise and bias in the objective function, which prevents reliable convergence. As a consequence and to draw a direct comparison, we use the full 2-design for the calculation of the infidelity for two qubits for both methods. Since performing the optimization with CRAB would be computationally prohibitive for the 3-qubit gate, we choose not to carry out this optimization, and choose an ensemble of $n_E=100$ randomly initialized states for the qualitative examination of the performance of the optimization with PEPRino. 
\begin{figure}
    \centering
    \includegraphics{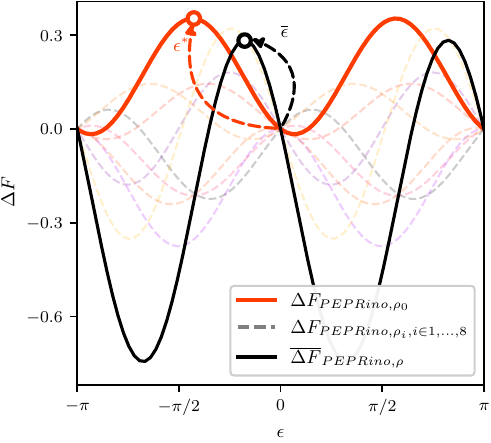}
    \caption{ \textbf{Change of the fidelity landscape for different initial states.} We show the change of the fidelity ${\Delta  F _{PEPRino}}$ as a function of $\epsilon$ for 8 random initial states $\rho_i$ (dashed lines), as well as the initial state $\rho_0$ (red line) which is used in Fig.~\ref{fig:Fig1}. ${\Delta F_{PEPRino}}$ is evaluated at the same random time $t_r$ and for the same control operator for all $n_B=9$ initial states $\rho$. This figure motivates that, for gate optimization, the averaged fidelity landscape has to be considered (black line) and an averaged update $\theta_{j,k}\rightarrow \theta_{j,k}-\bar{\epsilon} \frac{2}{t_f}\sin\left(\pi kt_r/t_f\right)$ is needed. An optimal update step for a random initial state can implement a low-fidelity update for another state, as shown for the optimal update (red arrow) for $\rho_0$ with the optimal value $\epsilon^*$. Thus, in gate optimization, a batch of initial states is used to determine averaged susceptibilities for the optimal averaged value $\bar{\epsilon}$ (black arrow).}
    \label{fig:Fig2}
\end{figure}

For state preparation operations, a single initial state is considered as a starting point, thus we apply the parameter update with $\epsilon =\epsilon^*$ obtained for the initial state. For gate optimization like the optimization of the implementation of the QFT, a batch of randomly sampled initial states $\rho_i(t_0)\equiv  \rho_i$ is used, and an averaged $\bar{\epsilon}$ is calculated from the averaged susceptibilities 
\begin{align}
    \overline{\chi^{(1,2)}}&=\frac{1}{n_B}\sum_{i=1}^{n_B}\chi^{(1,2)}_{\rho_i}.
    \label{eq:averagechi}
\end{align}
The average is performed over a batch of size $n_B$ and the update is performed with the resulting averaged parameter $\epsilon=\bar{\epsilon}$. This averaging is necessary due to the fact that the single update for a randomly sampled initial state is optimal for this specific state, but not in general, see Fig.~\ref{fig:Fig2} as an illustration. The fidelity landscape of a single initial state $\rho_0$, shown in red, differs from those corresponding to other initial states $\rho_i$, indicated by the dashed lines. Therefore, we generate a fidelity landscape averaged over a batch as in Eq.~\eqref{eq:averagechi}, as represented by the black line in Fig.~\ref{fig:Fig2}.

\section{Results}
\label{sec:high-fidelity-operations}
In the following, we show the results for the optimization with our proposed method in the 2- and 3-qubit setup. Additionally, the results of the optimization for two qubits are compared to the results obtained for the CRAB algorithm. For the initialization of both optimization algorithms, the transformation parameters $\{\theta\}_0$ of the pulses $h_{x/y}^j(t)$ and $J^j(t)$ are randomly sampled as $\theta_{x/y/J,j,k}\sim \mathcal{N}(0, 1/(n_m\sqrt{k}))$. Additionally, we define the four scalar parameters that define the standard Nelder-Mead method as coefficients
of reflection $\alpha=1$, expansion $\gamma = 2$, contraction $\sigma = 0.5$, and shrinkage $\beta = 0.5$~\cite{lagarias1998convergence}. The remaining vertices $\{\theta\}_i=\{\theta\}_0+\epsilon_v\hat{e_p}$ of the simplex are built using updates in all $p\in\{1,\cdots,N_\theta\}$ directions in the parameter space, with $\epsilon_v = 1$.
\begin{figure}
    \centering
    \includegraphics{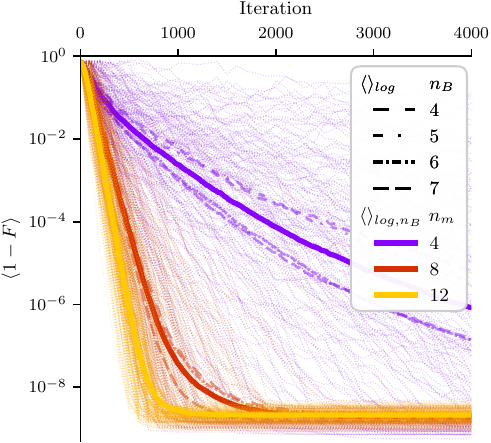}
    \caption{\textbf{Infidelity of the 2-qubit optimization process using PEPRino.} We show the infidelity during the optimization process of the QFT for two qubits as a function of iteration steps for PEPRino. Thin dotted lines represent individual optimization runs for different batch sizes $n_B\in\{4,5,6,7\}$ and numbers of modes $n_m\in\{4,8,12\}$ in the control pulse parameterization. The thick solid lines show the logarithmic average across all considered runs and batch sizes for the specified number of modes. Additionally, the thin solid lines with distinct line styles illustrate the logarithmic averages for each batch size,  with the mode number indicated by the corresponding color. Our method finds high-fidelity solutions for all considered setups, where we find that for increasing batch size, the optimization converges against the lower bound in a smaller amount of iteration steps, however, the behavior and overall picture remains qualitatively the same. For increasing number of modes, the number of iterations necessary for convergence decreases rapidly for small numbers, and slowly for higher mode numbers. We find that the convergence behavior for $n_m=8$ builds a suitable trade-off of necessary iteration steps and complexity in the calculations. \label{fig:Fig3}}
\end{figure}

We mainly focus on the high-fidelity results for the 2-qubit case, where a detailed analysis of the optimization performance is given. In addition, we present results for three qubits obtained using PEPRino to demonstrate the adaptability and scalability of our proposed method. For two qubits, we systematically investigate the dependence of the optimization performance on the number of modes in the control pulse parameterization, as well as on the batch size employed for the computation of the averaged susceptibilities. We further compare the results with those obtained via CRAB. Although both methods achieve comparable final infidelities, and thus precision, PEPRino exhibits a clear advantage in terms of computational efficiency, requiring less time and resources. Given the unfavorable scaling of CRAB with increasing qubit number and parameter space dimension, we restrict the 3-qubit analysis to those obtained with PEPRino, due to the computational cost of performing CRAB optimization.

\subsection{Optimal implementation of the 2-qubit-QFT}
\label{subsect:1t}
\begin{figure}
    \centering
    \includegraphics{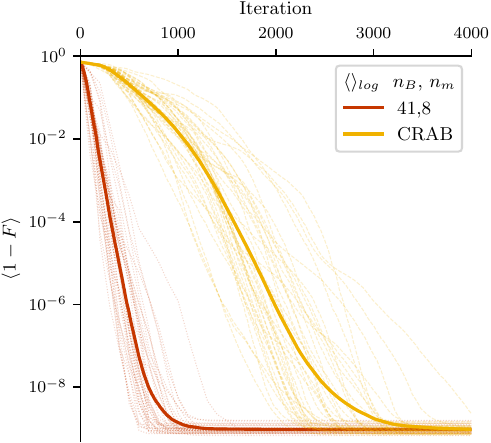}
    \caption{\textbf{PEPRino vs CRAB for the 2-qubit QFT optimization. }We show the infidelity obtained using PEPRino (red) and CRAB (orange) as a function of iterations. For PEPRino, we choose $n_m = 8$ with a batch size of $n_B=41$, in accordance with the size of the simplex used in CRAB, where we also consider $n_m = 8$. The dashed lines are individual optimization runs, and the solid lines are the logarithmic average over the runs. Both methods converge against the same lower bound, which is reached in fewer iterations utilizing our proposed method. Additionally, the variance of the individual runs and computational time for individual iteration steps are less for PEPRino compared to CRAB. \label{fig:Fig4}}
\end{figure}
For the 2-qubit system, the control operators are given by expression~(\ref{eq:2qubitoperators}). We consider a parameterization of the time-dependent control pulses in terms of $n_m = 8$ modes for the main results, and compare this choice with $n_m = 4$ and $n_m =12$. The optimization results achieved with PEPRino are shown in Fig.~\ref{fig:Fig3}. The thin solid lines with distinct line styles represent the logarithmic average of 40 optimization runs for the indicated batch sizes. Here, the colors indicate the number of modes, which are purple, red and orange for $n_m=4$, $n_m=8$, and $n_m=12$, respectively. The thicker solid lines correspond to the overall logarithmic average across all considered batch sizes for the corresponding number of modes. We observe that increasing the number of modes generally reduces the number of iterations required for convergence. For instance, for $n_m=12$, convergence to high-fidelity implementations of the quantum Fourier transform (QFT) can be achieved in as few as $n_{it}\approx300$ iterations. However, the increased dimensionality of the parameter space leads to a higher computational cost per iteration, resulting in an overall slowdown. We therefore identify $n_m=8$ as a suitable trade-off between convergence speed and computational efficiency. For the computation of the average susceptibilities, we employ batch sizes in the range $n_B\in \{4,5,6,7\}$. We find that larger batch sizes generally lead to a reduction in the number of iterations required for convergence while yielding qualitatively similar lower bounds for the achieved infidelity. The infidelity is evaluated for and averaged over an ensemble of $n_E = 36$ initial states, which we choose to be the full 2-design for two qubits, see App.~\ref{app:sec-2design}, thereby providing an accurate estimate of the average state fidelity. 

To draw a performance comparison with CRAB, we optimize the gate implementation using this method, with the results shown in Fig.~\ref{fig:Fig4}. Here, we focus on the case $n_m=8$. The orange solid line shows the optimal results, obtained with CRAB, logarithmically averaged over an ensemble of 40 optimization runs. These are compared to results obtained with PEPRino, using a batch size of $n_B=41$ (red lines). This choice is motivated by the structure of the CRAB simplex, which consists of $N_v=(3n_q-1)n_m+1$ vertices, yielding $N_v = 41$ for the parameters considered. Accordingly, we set the batch size in PEPRino to the same value for a fair comparison. With this choice, both methods converge to a similar lower bound in infidelity. However, PEPRino requires significantly fewer iteration steps ($n_{it,PEPRino}\approx 1000$) compared to CRAB ($n_{it,CRAB}\approx 3000$), while also demanding fewer computational resources. Notably, for PEPRino, a batch size of $n_B=8$ is already sufficient to achieve comparable results, albeit with a slightly higher infidelity than in the $n_B=41$ case. Overall, both batch sizes yield satisfactory performance, with smaller batch sizes offering the advantage of reduced computational cost. Larger batch sizes are therefore not strictly necessary, but may be beneficial for final fine-tuning in the optimization. 

\subsection{Optimal implementation of the 3-qubit-QFT}\label{subsect:2}
\begin{figure}
    \centering
    \includegraphics{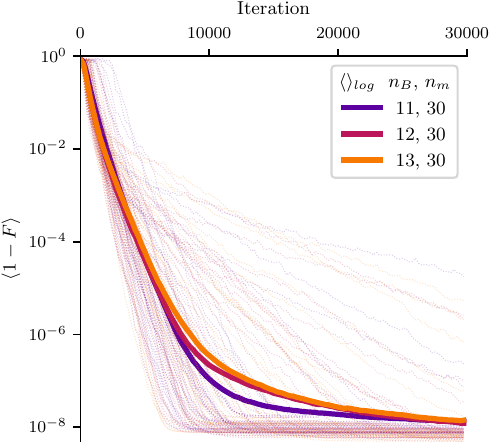}
    \caption{\textbf{Infidelity of the 3-qubit optimization process using PEPRino.} We present the infidelity during the optimization process of the QFT for three qubits as a function of iteration steps for PEPRino. For three qubits, we consider $n_m=30$ modes in the control pulse parameterization. We show the results for batch sizes $n_B=\{11,12,13\}$. The solid lines are the logarithmic average over an ensemble of 40 optimization runs for the given batch sizes. The findings agree with the findings for the 2-Qubit-QFT optimization. For increasing batch sizes, we find a tendency towards convergence in fewer iteration steps, however, the behavior and result remain generally similar for the range of batch sizes used here. We find optimal solutions of the implementation with ${n_{it}\approx 6000}$ iteration steps.\label{fig:Fig5}}
\end{figure}
As a second example, we consider the optimization of the 3-qubit-QFT. For $n_q=3$ qubits, the implementation becomes more complex, we therefore parameterize the time-dependent control pulses with $n_m=30$ modes and use batch sizes $n_B=\{11,12,13\}$. For both optimization methods, increasing the dimensionality of the parameter space leads to longer optimization times, however, for CRAB this effect is significantly more pronounced. In the CRAB approach, each update step is performed using a simplex consisting of $N_v=241$ vertices. This approach requires evaluation of the performance at each vertex, as well as at additional candidate points generated during the update. Each such evaluation involves computing the averaged fidelity over the 3-qubit 2-design, which, following our scheme for the formation of 2-designs for $N$ qubits, see App.~\ref{app:sec-2design}, requires 216 time evolutions. As a result, the total number of time evolutions required for a single successful update step, and thus for overall successful optimization, quickly exceeds what can be considered rational computational cost. In addition, as already demonstrated for the 2-qubit case, PEPRino outperforms CRAB in terms of convergence behavior and computational efficiency. Therefore, we do not pursue the CRAB method for three qubits. For PEPRino, we show the corresponding results in Fig.~\ref{fig:Fig5}. We present the average infidelity of an ensemble of $n_E=100$ randomly initialized states to visualize the gate infidelity for individual runs and logarithmically averaged for an ensemble of 40 runs for each batch size. While both the number of iterations required for convergence and the computational effort per iteration increase compared to the 2-qubit case, the method remains successful. We find convergence towards a lower bound in the infidelity for $n_{it}>6000$, demonstrating that high-fidelity implementations of the 3-qubit-QFT can be achieved efficiently within this framework. 

\section{Conclusion}
\label{sec:conclusion}
In conclusion, we put forth a gradient-free and hyperparameter-free optimization algorithm for the control of quantum processes. For a given system that contains control terms composed of control functions and control operators, the algorithm maps out the environment of the fidelity landscape, by resumming an infinite series of response functions into a manageable expression. Based on this, the algorithm identifies an optimal update step. By iterating this step, the algorithm converges to an optimal implementation of a target operation.

We have demonstrated the performance of the algorithm on proof-of-concept examples of 2- and 3-qubit systems. In the 2-qubit setting, a direct comparison with the established optimization method CRAB using the Nelder-Mead algorithm showed that our method provides improved convergence speed and stability with decreased variance in the individual optimization runs, resulting in high-fidelity implementations of the desired transformation. The extension to three qubits further highlighted the scalability and robustness of the approach in higher-dimensional settings.

Overall, the results indicate that the proposed framework constitutes a promising alternative for quantum optimal control tasks, particularly in scenarios where traditional methods suffer from sensitivity to initialization and hyperparameter tuning. Its flexibility makes it applicable to a wide range of problems, including quantum gate optimization and state preparation, with improved usability due to its hyperparameter-free manner.

Given that any implementation of quantum technologies under realistic conditions will naturally entail optimization and optimal control of quantum processes, the algorithm put forth here supports the development of quantum technologies by advancing the necessary optimization algorithms.

\begin{acknowledgments}
This work is funded by the Cluster of Excellence 'CUI: Advanced Imaging of Matter' of the Deutsche Forschungsgemeinschaft (DFG) (EXC 2056), project ID 390715994. The project is co-financed by ERDF of the European Union and by ’Fonds of the Hamburg Ministry of Science, Research, Equalities and Districts (BWFGB)’.
\end{acknowledgments} 

\bibliography{main.bib}

\onecolumngrid
\appendix
\section{Nested Pauli commutators for two qubits}  
\label{app:sec-Paulimatrices}
The 2-qubit density matrix takes the form

\begin{equation}
\rho = \frac{1}{4} \sum_{a=0}^{3} \sum_{b=0}^{3} \rho_{a,b} \, \sigma_a \otimes \sigma_b.
\end{equation}

We show with the example of the control operator
\begin{equation}
B= \sigma_z \otimes \sigma_z =
\begin{pmatrix}
1 & 0 & 0 & 0 \\
0 & -1 & 0 & 0 \\
0 & 0 & -1 & 0 \\
0 & 0 & 0 & 1
\end{pmatrix}
\end{equation}
that we recover the proposed relation Eq.~(\ref{eq:nestedPaulicommutatorodd}) and Eq.~(\ref{eq:nestedPaulicommutatoreven}) of higher order commutators being proportional to the first and second order commutator of the density matrix and the control operator, for control operators in the form of products of Pauli matrices. The first order commutator is given by
\allowdisplaybreaks
\begin{align}
    i\left[B,\rho\right] &= \frac{i}{4}\left[\sigma_3\otimes\sigma_3, \sum_{a=0}^{3} \sum_{b=0}^{3} \rho_{a,b} \, \sigma_a \otimes \sigma_b\right] \\
    &= \frac{i}{4} \sum_{a=0}^{3} \sum_{b=0}^{3} \rho_{a,b}\left[\sigma_3\otimes\sigma_3, \, \sigma_a \otimes \sigma_b\right] \\
    &= \frac{i}{4} \sum_{a=0}^{3} \sum_{b=0}^{3} \rho_{a,b}\left(\left[\sigma_3,\sigma_a\right]\otimes \sigma_3\sigma_b+\sigma_a\sigma_3 \otimes\left[\sigma_3,\sigma_b\right]\right)  \\
    &= \frac{i}{4} \sum_{a=1}^{2} \sum_{b=0}^{3} \rho_{a,b}\left( 2i \sum_k\varepsilon_{3ak} \sigma_k\otimes \sigma_3\sigma_b\right)+\frac{i}{4} \sum_{a=0}^{3} \sum_{b=1}^{2} \rho_{a,b}\left(2i\sum_l\sigma_a\sigma_3 \otimes\varepsilon_{3bl} \sigma_l\right)  \\
    &= - \frac12 \sum_{a=1}^{2} \sum_{b=0}^{3} \rho_{a,b}\left(\sum_k \varepsilon_{3ak} \sigma_k\otimes \left(\delta_{0b}\sigma_3+\delta_{3b}\mathds{1}+\sum_m i\epsilon_{3bm}\sigma_m\right)\right)
    +\\
    &\hspace{1cm}-\frac12\sum_{a=0}^{3} \sum_{b=1}^{2} \rho_{a,b}\left(\left(\delta_{0a}\sigma_3+\delta_{3a}\mathds{1}+\sum_ni\epsilon_{a3n}\sigma_n\right)\otimes\, \sum_l\varepsilon_{3bl} \sigma_l\right)  \\
    &= - \frac{i}{2}\sum_{a=1}^{2} \sum_{b=1}^{2} \rho_{a,b} \varepsilon_{3ak}\epsilon_{3bm} \sigma_k\otimes \sigma_m -\frac12\sum_{a=1}^{2}\rho_{a,3}\varepsilon_{3ak} \sigma_k\otimes \mathds{1}-\frac12\sum_{a=1}^{2}\rho_{a,0}\varepsilon_{3ak} \sigma_k\otimes \sigma_3+\\
    &\hspace{1cm}-\frac{i}{2}\sum_{a=1}^{2} \sum_{b=1}^{2}\rho_{a,b}\epsilon_{a3n} \varepsilon_{3bl}\sigma_n\otimes\, \sigma_l-\frac12 \sum_{b=1}^{2} \rho_{3,b}\varepsilon_{3bl}\mathds{1}\otimes\,  \sigma_l-\frac12\sum_{b=1}^{2} \rho_{0,b}\varepsilon_{3bl}\sigma_3\otimes\, \sigma_l\\
    &=-\frac12\sum_{a=1}^{2}\rho_{a,3}\varepsilon_{3ak} \sigma_k\otimes \mathds{1}-\frac12\sum_{a=1}^{2}\rho_{a,0}\varepsilon_{3ak} \sigma_k\otimes \sigma_3-\frac12 \sum_{b=1}^{2} \rho_{3,b}\varepsilon_{3bl}\mathds{1}\otimes\,  \sigma_l-\frac12\sum_{b=1}^{2} \rho_{0,b}\varepsilon_{3bl}\sigma_3\otimes\, \sigma_l.
\end{align}
For the second order commutator, we find
\begin{align}
    i^2\left[B,\left[B,\rho\right]\right] &= i^2\left[B,\left[B,\frac{1}{4} \sum_{a=0}^{3} \sum_{b=0}^{3} \rho_{a,b} \, \sigma_a \otimes \sigma_b\right]\right]\\
    &=i\left[\sigma_3\otimes\sigma_3,-\frac12\sum_{a=1}^{2}\rho_{a,3}\varepsilon_{3ak} \sigma_k\otimes \mathds{1}\right] + i\left[\sigma_3\otimes\sigma_3,-\frac12\sum_{a=1}^{2}\rho_{a,0}\varepsilon_{3ak} \sigma_k\otimes \sigma_3\right]+\\
    &\hspace{1cm}i\left[\sigma_3\otimes\sigma_3,-\frac12 \sum_{b=1}^{2} \rho_{3,b}\varepsilon_{3bl}\mathds{1}\otimes\,  \sigma_l\right]+i\left[\sigma_3\otimes\sigma_3,-\frac12\sum_{b=1}^{2} \rho_{0,b}\varepsilon_{3bl}\sigma_3\otimes\, \sigma_l\right].
\end{align}
Evaluating all four terms results in 
\begin{align}
    i^2\left[B,\left[B,\rho\right]\right]& = \sum_{a=1}^{2}\rho_{a,3}\varepsilon_{3ak}\varepsilon_{3kl} \sigma_l\otimes \sigma_3 +\sum_{a=1}^{2}\rho_{a,0}\varepsilon_{3ak}\varepsilon_{3kl} \sigma_l\otimes \mathds{1}+\sum_{b=1}^{2} \rho_{3,b}\varepsilon_{3bl}\varepsilon_{3lk} \sigma_3 \otimes\sigma_k+\sum_{b=1}^{2} \rho_{0,b}\varepsilon_{3bl}\varepsilon_{3lk} \mathds{1}\otimes \sigma_k.
\end{align}
Analogously, the third order commutator can be computed, and we find for this expression
\begin{align}
    i^3\left[B,\left[B,\left[B,\rho\right]\right]\right]& =2\sum_{a=1}^{2}\rho_{a,3}\varepsilon_{3ay} \sigma_y\otimes\mathds{1}+2\sum_{a=1}^{2}\rho_{a,0}\varepsilon_{3ay} \sigma_y\otimes\sigma_3++2\sum_{b=1}^{2} \rho_{3,b}\varepsilon_{3by}\mathds{1}\otimes\sigma_y+2\sum_{b=1}^{2} \rho_{0,b}\varepsilon_{3by}\sigma_3\otimes\sigma_y.
\end{align}
We can now compare the results for the first to third order commutator of the control operator with the general density matrix:
\begin{align}
    i\left[B,\rho\right] &=-\frac12\sum_{a=1}^{2}\rho_{a,3}\varepsilon_{3ak} \sigma_k\otimes \mathds{1}-\frac12\sum_{a=1}^{2}\rho_{a,0}\varepsilon_{3ak} \sigma_k\otimes \sigma_3-\frac12 \sum_{b=1}^{2} \rho_{3,b}\varepsilon_{3bl}\mathds{1}\otimes\,  \sigma_l-\frac12\sum_{b=1}^{2} \rho_{0,b}\varepsilon_{3bl}\sigma_3\otimes\, \sigma_l \\ 
    i^2\left[B,\left[B,\rho\right]\right]& = \sum_{a=1}^{2}\rho_{a,3}\varepsilon_{3ak}\varepsilon_{3kl} \sigma_l\otimes \sigma_3 +\sum_{a=1}^{2}\rho_{a,0}\varepsilon_{3ak}\varepsilon_{3kl} \sigma_l\otimes \mathds{1}+\sum_{b=1}^{2} \rho_{3,b}\varepsilon_{3bl}\varepsilon_{3lk} \sigma_3 \otimes\sigma_k+\sum_{b=1}^{2} \rho_{0,b}\varepsilon_{3bl}\varepsilon_{3lk} \mathds{1}\otimes \sigma_k\\
    i^3\left[B,\left[B,\left[B,\rho\right]\right]\right]& =2\sum_{a=1}^{2}\rho_{a,3}\varepsilon_{3ay} \sigma_y\otimes\mathds{1}+2\sum_{a=1}^{2}\rho_{a,0}\varepsilon_{3ay} \sigma_y\otimes\sigma_3+2\sum_{b=1}^{2} \rho_{3,b}\varepsilon_{3by}\mathds{1}\otimes\sigma_y+2\sum_{b=1}^{2} \rho_{0,b}\varepsilon_{3by}\sigma_3\otimes\sigma_y\\
    &\hspace{1cm} = -4\cdot i\left[B,\rho\right].
\end{align}
Thus, we recover Eq.~(\ref{eq:nestedPaulicommutatorodd}) and Eq.~(\ref{eq:nestedPaulicommutatoreven}) not only for single-qubit operators, but also for products of Pauli matrices as 2-qubit control operators. This pattern extends to $N$ qubits and can be used in the calculation of higher order susceptibilities, i.e.
 \begin{align}
\chi^{(n)}(t_{r}) &= \left(\frac{i}{\hbar}\right)^n Tr\Big(\rho_\text{target}U(t_f,t_r)[B,[\ldots[,[B,U(t_r)\rho(t_0)U^\dagger(t_r)]\ldots]_nU^\dagger(t_f,t_r)\Big).
\end{align}
\allowdisplaybreaks[0]

\section{Detailed derivation of the method}
\label{app:sec-derivation-series}
We consider the Hamiltonian $H_0$ of the unperturbed system, as well as a perturbation of the form 
\begin{align*}
    H_{\text{pr}}(t) = -F(t)B.
\end{align*}
The response of the system to this perturbation is discussed. The general form of the mean value of an observable $A$ at a time $t$ in the Heisenberg picture is
\begin{align}
    \langle A(t)\rangle &=\Tr{\left(\rho(t)A\right)}= \Tr{\left(U(t,t_0)\rho(t_0)U^\dagger (t,t_0)A\right)}.
    \label{appeq:expectation_A_ansatz}
\end{align}
Here, $U(t,t_0)$ is the time evolution operator. The Schroedinger equation
\begin{align*}
    i\hbar \frac{\partial}{\partial t}|\psi(t)\rangle = H|\psi(t)\rangle
\end{align*}
is formally solved by
\begin{align*}
    |\psi(t)\rangle = U_0(t,t_0)|\psi(t_0)\rangle
\end{align*}
with
\begin{align*}
    U_0(t,t_0)=e^{-iH(t-t_0)/\hbar}.
\end{align*}
The Heisenberg picture is formulated as
\begin{align*}
    |\psi_H\rangle = U_0^\dagger(t,t_0)|\psi(t)\rangle = |\psi(t_0)\rangle,~A(t) = U_0^\dagger(t,t_0)AU_0(t,t_0),
\end{align*}
with the Heisenberg equation of motion being
\begin{align*}
    i\hbar \frac{\partial}{\partial t}A(t) = \frac{i}{\hbar}[H,A(t)]
\end{align*}
Using the characteristic of cyclic invariance of the trace, Eq.~(\ref{appeq:expectation_A_ansatz}) corresponds to 
\begin{align}
    \langle A(t)\rangle &=\Tr{\left(\rho(t_0)U^\dagger (t,t_0)AU(t,t_0)\right)}.
    \label{appeq:expectation_A_secondansatz} 
\end{align}
Taking a closer look at the time evolution operator, we start working in the interaction picture. To do so, we first evaluate the equation of motion for the time evolution operator
\begin{align*}
    i\hbar \frac{\partial}{\partial t}|\psi(t)\rangle &= i\hbar \frac{\partial}{\partial t}U(t,t_0)|\psi(t_0)\rangle  = H|\psi(t)\rangle = H U(t,t_0)|\psi(t_0)\rangle\\
    \Rightarrow~& i\hbar \frac{\partial}{\partial t}U(t,t_0) = H U(t,t_0).
\end{align*}
For the time evolution operator and Hamiltonian in the interaction picture, this yields
\begin{align}    
H_{\text{pr,I}}(t)&=e^{iH_0(t-t_0)/\hbar}H_{\text{pr}}e^{-iH_0(t-t_0)/\hbar},\\
    U_I(t,t_0)&=  1 + \frac{1}{i\hbar}\int_{t_0}^t   dt'H_{\text{pr,I}}(t')U_I(t',t_0)=\\
    &=1 + \frac{1}{i\hbar}\int_{t_0}^t   dt'H_{\text{pr,I}}(t')+\frac{1}{(i\hbar)^2}\int_{t_0}^t   dt'\int_{t_0}^{t'}   dt''H_{\text{pr,I}}(t')H_{\text{pr,I}}(t'')+\ldots
\end{align}
We can continue working on Eq.~(\ref{appeq:expectation_A_secondansatz}) with this expression:
\begin{align}
    \langle A(t)\rangle &=\Tr\left(\rho(t_0)U^\dagger (t,t_0)A_IU(t,t_0)\right)\\
    &=\Tr\Bigg( \rho(t_0)\Big( 1 - \frac{1}{i\hbar}\int_{t_0}^t   dt'H_{\text{pr,I}}(t')+\frac{1}{(i\hbar)^2}\int_{t_0}^t   dt'\int_{t_0}^{t'}   dt''H_{\text{pr,I}}(t')H_{\text{pr,I}}(t'')-\ldots\Big)\times \nonumber\\
    &\quad\quad \times  A_I\times\Big( 1 + \frac{1}{i\hbar}\int_{t_0}^t   dt'H_{\text{pr,I}}(t')+\frac{1}{(i\hbar)^2}\int_{t_0}^t   dt'\int_{t_0}^{t'}   dt''H_{\text{pr,I}}(t')H_{\text{pr,I}}(t'')+\ldots\Big)\Bigg)
    \label{appeq:expectation_of_A_full}
\end{align}
In 0-th order, this gives the expectation value of $A$ in the unperturbed system, i.e.
\begin{align}
    \langle A(t)\rangle = \Tr(\rho(t_0)A(t)) = \langle A \rangle_0.
\end{align}
To 1st order, Eq.~\eqref{appeq:expectation_of_A_full} gives
\begin{align}
    \Delta \langle A(t)\rangle_1 &=\langle A(t)\rangle -\langle A(t) \rangle_0\\
    &=\Tr\Big( -\rho(t_0)\frac{1}{i\hbar}\int_{t_0}^t   dt'H_{\text{pr,I}}(t')A(t)+\rho(t_0)A(t) \frac{1}{i\hbar}\int_{t_0}^t   dt'H_{\text{pr,I}}(t')\Big)\\
    &=\frac{1}{i\hbar} \int_{t_0}^t   dt'\langle[A(t),H_\text{pr,I}(t')]\rangle_0
\end{align}
To 2nd order:
\begin{align}
    \Delta \langle  &A(t)\rangle_2=\langle A(t)\rangle -\langle A(t) \rangle_0-\frac{1}{i\hbar} \int_{t_0}^t   dt'\langle[A(t),H_\text{pr,I}(t')]\rangle_0\\
    &=\Tr\Bigg(\rho(t_0)\Big(A(t)\frac{1}{(i\hbar)^2}\int_{t_0}^t   dt'\int_{t_0}^{t'}   dt''H_{\text{pr,I}}(t')H_{\text{pr,I}}(t'')-\frac{1}{i\hbar}\int_{t_0}^t   dt'H_{\text{pr,I}}(t')A(t)\frac{1}{i\hbar}\int_{t_0}^t   dt''H_{\text{pr,I}}(t'') +\nonumber\\
    &\quad\quad\quad\quad+\frac{1}{(i\hbar)^2}\int_{t_0}^t   dt'\int_{t_0}^{t'}   dt''H_{\text{pr,I}}(t')H_{\text{pr,I}}(t'')A(t)\Big)\Bigg) \\
    &=\frac12\frac{1}{(i\hbar)^2} \int_{t_0}^t   dt'\int_{t_0}^{t}   dt''\mathcal{T}\langle[[A(t),H_\text{pr,I}(t')],H_\text{pr,I}(t'')]\rangle_0
\end{align}
To n-th order:
\begin{align}
    \Delta \langle A(t)\rangle_n &=\frac{1}{n!}\frac{1}{(i\hbar)^n} \int_{t_0}^t   dt'\ldots dt'^{n}\mathcal{T}\langle[\ldots[[A(t),H_\text{pr,I}(t')],H_\text{pr,I}(t'')],\ldots,H_\text{pr,I}(t'^{n})]_n\rangle_0 \\
    \rightarrow\Delta \langle A(t)\rangle &= \langle A(t)\rangle -\langle A \rangle_0=\nonumber\\
    &=\sum_n\frac{1}{n!}\frac{1}{(i\hbar)^n} \int_{t_0}^t   dt'\ldots dt'^{n}\mathcal{T}\langle[\ldots[[A(t),H_\text{pr,I}(t')],H_\text{pr,I}(t'')],\ldots,H_\text{pr,I}(t'^{n})]_n\rangle_0 .
    \label{appeq:seriesexpansion_expectation_A}
\end{align}
Now, considering the perturbation time local at random time $t_r$, i.e.
\begin{align}
    H_\text{pr,I}=-\epsilon\delta(t-t_r)B = -F(t)B,
\end{align}
Eq.~(\ref{appeq:seriesexpansion_expectation_A}) simplifies to
\begin{align}
    \Delta \langle A(t)\rangle&=\sum_n\frac{(-\epsilon)^n}{n!}\frac{1}{(i\hbar)^n} \langle[\ldots[[A(t),B(t_r)],B(t_r)],\ldots,B(t_r)]\rangle_0 .
\end{align}
The discussed observable $A$ is set to $\rho_\text{target}$. Then, one finds that
\begin{align}
    \Delta \langle A(t)\rangle=\Delta F_{\theta}&=
 \sum_{n=1}^{\infty}\Big(\frac{i\epsilon}{\hbar}\Big)^{n} \frac{1}{n!} \Tr\Big( [ \ldots ,[ U^{\dagger}(t_{f}) \rho_\text{target} U(t_{f}), U^{\dagger}(t_{r}) B U(t_{r})], \ldots ] \rho(0) \Big)\\
 &= \sum_{n=1}^{\infty} \frac{1}{n!} \epsilon^{n} \chi^{(n)}(t_{r})
    \end{align}
with
 \begin{align}
\chi^{(n)}(t_{r}) &= \Big(\frac{i}{\hbar}\Big)^{n}  
\Tr\Big( \rho_\text{target} U(t_{r}, t_{f}) [ B, \ldots[B,  U(t_{r}) \rho(0) U^\dagger(t_{r}) ]\ldots ]  U^{\dagger}(t_{r}, t_{f})  \Big)
    \end{align}
Using the series expression of sine and cosine, we consider the function \(\Delta  F_\theta \) expanded as:
\begin{align}
\Delta F_\theta &=
\chi^{(1)}\left(\epsilon+(-4)\epsilon^3/3!+(-4)^2\epsilon^5/5!+\ldots\right)+\chi^{(2)}\left(\epsilon^2/2+(-4)\epsilon^4/4!+(-4)^2\epsilon^6/6!+\ldots\right)\\
&=\frac{1}{2} \chi^{(1)} \sin(2\epsilon) + \frac{1}{4} \chi^{(2)} \left(1 - \cos(2\epsilon) \right),
\end{align}
where \(\chi^{(1)}\) and \(\chi^{(2)}\) are the first and second-order susceptibility. 
To find the value of \(\epsilon\) that maximizes \(\Delta F_\theta\), we compute the critical points by solving:
\[
\frac{d}{d\epsilon} \Delta F_\theta= \chi^{(1)} \cos(2\epsilon) + \frac12 \chi^{(2)}  \sin(2\epsilon) = 0.
\]
This gives the critical points
\[
\epsilon_n = \frac{1}{2} \arctan\left(\frac{-2\chi^{(1)} }{\chi^{(2)}}\right) + \frac{n\pi}{2}, \quad n \in \mathbb{Z}.
\]
We evaluate the second derivative to identify the extrema and select among the $\epsilon_n$ resulting in maxima the \(\epsilon^*\) that is smallest in absolute value.
The parameter updates for the trainable parameters $\theta_{j,k}$ then are
\begin{align*}
\theta_{j,k}\rightarrow \theta_{j,k}-\epsilon^* \frac{2}{t_f}\sin\left(\pi kt_r/t_f\right).
\end{align*} 

\section{2-designs}  
\label{app:sec-2design}
For a single qubit, a convenient starting point for constructing a 2-design~\cite{PhysRevA.80.012304} is the Clifford group $C_1$, whose elements can be written as $\mathbf{A}\mathbf{B}$, where $\mathbf{A} \in \{\mathds{1}, H, S, HS, SH, HSH\}$ and $\mathbf{B} \in \{\mathds{1}, X, Y, Z\}$. This construction yields 24 distinct elements $\mathbf{A}\mathbf{B}$. However, when applied to the initial state $\rho_0 = \tfrac{1}{2}(\mathds{1} + Z) = \ket{0}$, only 6 distinct states are generated in the resulting 2-design $S$ (up to a global phase).
\begin{equation}
    \textbf{AB}\ket{0}\in\{\ket{0},\ket{1},\ket{+},\ket{-},\ket{+ i},\ket{-i}\}
\end{equation}
where $\ket{\pm i} = \frac{1}{\sqrt2}\left(\ket{0}\pm i\ket{1}\right)$. The 6 projectors are $P_0 = |0\rangle\langle 0| = \tfrac{1}{2}(\mathds{1} + Z),~ P_1 = \tfrac{1}{2}(\mathds{1} - Z),~P_{+} = \tfrac{1}{2}(\mathds{1} + X),~ P_{-} = \tfrac{1}{2}(\mathds{1} - X),~P_{+i} = \tfrac{1}{2}(\mathds{1} + Y)$ and $P_{-i} = \tfrac{1}{2}(\mathds{1} - Y)$. So, the general density matrix can be expressed in terms of the projectors $P_{\psi}$:
\begin{equation}
\rho = \tfrac{1}{2}\left( \mathds{1} + \rho_x X + \rho_y Y + \rho_z Z \right)
= \tfrac{1}{6} \sum_{\psi \in S} P_{\psi}
+ \tfrac{\rho_x}{2}(P_+ - P_-)
+ \tfrac{\rho_y}{2}(P_{+i} - P_{-i})
+ \tfrac{\rho_z}{2}(P_0 - P_1).
\end{equation}
We can show that the set actually covers the entire density matrix: \[\tfrac{1}{6}\sum_{\psi \in S} P_{\psi} 
{=} \int \rho(\theta,\phi)\, d\mu(\theta,\phi)
\]
using
\[
\int d\mu(\theta,\phi) = 1
\quad \text{and}\quad 
\int_0^{2\pi}\!\!\int_0^{\pi} \sin\theta\, d\theta\, d\phi = 4\pi.
\]
To form the $N$-qubit 2-design, we use the 1-Qubit 2-design $S$, consisting of 6 quantum states, as a $N$-tensor product with itself, resulting in $6^N$ quantum states in the $N$-qubit 2-design. 
\begin{figure}
    \centering
    \includegraphics{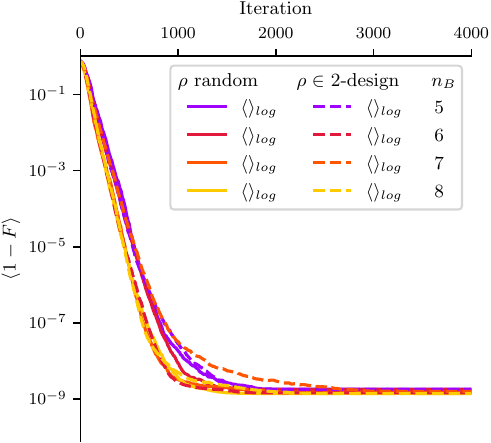}
    \caption{\textbf{Random sampling of $\rho(t_0)$ vs sampling from the 2-qubit 2-design.} We show the averaged infidelity for the 2-qubit-QFT optimization over 20 runs as a function of iterations steps, for the corresponding batch sizes and $n_m=8$ modes in the parameterization of the control pulses. We distinguish between randomly initialized density matrices (solid lines) and density matrices sampled from the 2-qubit 2-design (dashed lines), which are used to calculate the averaged susceptibilities for the parameter update. Both sampling methods lead to comparable results and are therefore equally applicable. \label{fig:Fig_6}}
\end{figure}

For $N$ qubits, one way to sample initial states is by stochastic sampling, where we sample $N\times 3$ real numbers $\rho_{x,i}, ~\rho_{y,i}, ~\rho_{z,i}$ and normalize the Bloch vector to ensure a valid quantum state, then construct the individual density matrix as $\rho_i = \tfrac{1}{2} \left( \mathds{1}+ \rho_{x,i} X + \rho_{y,i} Y + \rho_{z,i} Z \right)$ and the total density matrix as the tensor product of the individual density matrices. Alternatively, we can sample states from the $N$-qubit 2-design. This is shown exemplary for two qubits in Fig.~\ref{fig:Fig_6}, where we show the result of the optimization of the QFT with $n_m=8$ modes in the control pulse parameterization as a function of iteration steps for a range of batch sizes. The batch size is used to calculate the average susceptibilities, where the density matrices are either randomly sampled (solid lines) or drawn from the 2-design (dashed lines). We find that for the outcome of the optimization process, both ways result in similar results. Thus, for the sampling process, both methods are valid. For the evaluation of the infidelity, the full 2-design is used in order to capture the full randomization.

\end{document}